# The role of surface depletion layer effects on the enhancement of the UV emission in ZnO induced by a nanostructured Al surface coating.

*Saskia Fiedler\*[1,2], Laurent O. Lee Cheong Lem[1,3], Cuong Ton-That[1], Matthew R. Phillips\*[1]*

Dr. Saskia Fiedler, Dr. Laurent O. Lee Cheong Lem, Assoc. Prof. Cuong Ton-That, Prof. Matthew R. Phillips

1. School of Mathematical and Physical Sciences, University of Technology Sydney, 15 Broadway, Ultimo NSW 2007, Australia

Dr. Saskia Fiedler

2. Centre for Nano Optics, University of Southern Denmark, Campusvej 55, 5230 Odense M, Denmark

Dr. Laurent O. Lee Cheong Lem

3. Australian National Fabrication Facility, Australian National University, Canberra ACT 2601, Australia

E-mail: safi@mci.sdu.dk and matthew.phillips@uts.edu.au

## Abstract

The UV enhancement of Al-coated ZnO single crystals with a wide range of carrier densities is systematically studied using depth-resolved cathodoluminescence (CL) and photoluminescence (PL) as well as valence band X-ray photoemission spectroscopy (VB-XPS). An up to 17-fold enhanced PL UV emission for Al-coated ZnO with the highest carrier density ($2 \times 10^{17}$ cm$^{-3}$) was measured, which falls to a 12-fold increase for the lowest carrier density ($3 \times 10^{13}$ cm$^{-3}$). Depth-resolved CL measurements confirm that the enhancement is strongest near the metal coating – ZnO interface consistent with an increased UV emission due to an exciton – localised surface plasmon coupling mechanism. Correlative CL, PL and VB-XPS studies reveal that a number of additional effects related to the presence of the Al surface coating also contribute to the UV enhancement factor. These include increased UV enhancement due to the formation of a surface depletion layer induced by the metal surface coating, which also passivates competitive non-radiative surface recombination channels found in uncoated ZnO. Significantly, it was established that the magnitude of the emission enhancement factor can be raised in a controlled way by reducing the thickness of the depletion layer by increasing the carrier density. The contribution of these effects collectively provides an explanation for the large span of enhancement factors reported in the literature.

**Introduction**

Metal nanoparticle (NP) surface coatings can significantly enhance the excitonic luminescence of zinc oxide (ZnO) and other semiconductors [1–5] currently used in a broad array of applications in optoelectronics, photonics, photovoltaics and energy technologies. Here, a direct dipole-dipole coupling between the localized surface plasmons (LSPs) in the metal nanoparticles and the excitons in the ZnO creates an additional, non-radiative, faster relaxation channel via the LSPs, which subsequently leads to an increased spontaneous emission rate of the UV excitonic near band edge (NBE) emission [5,6]. This coupling mechanism is illustrated in fig. 1.

Gold (Au) and silver (Ag) are the most commonly used metal NPs for plasmonic applications. Although, Au NP coatings are relatively easy to fabricate as they are inert and stable, their plasmonic resonance is in the green spectral range, which makes Au films unsuitable for plasmonic coupling to the UV-excitons in ZnO [1,7,8]. Ag NPs, on the other hand, have their plasmon resonance in the blue/UV region; however, they readily form oxide and sulfur layers, which makes the fabrication of stable Ag NPs films difficult [9,10]. Conversely, Al NP films have a widely tunable plasmon resonance energy extending from the UV to the visible spectrum, depending on the size and the shape of the NPs [11–14]. Even though Al NPs easily oxidize when exposed to air, the oxide layer thickness is typically self-limiting, which can be used to control the plasmon resonance energy [15]. Indeed, the UV excitonic emission in ZnO has been demonstrated to increase when the sample is coated with a surface film of Al NPs, which has been widely attributed to an mechanism involving exciton – LSP coupling. However, enhancement factors extending over a few orders of magnitude have been reported from samples that seem to be identical by their description [3,4,16–20], suggesting that there are additional effects contributing to the measured Al metal film induced increase in the light emission.

In this work, the UV enhancement of Al-coated ZnO single crystals with a wide range of carrier densities was systematically studied using photoluminescence (PL) spectroscopy and depth-resolved cathodoluminescence (CL). An up to 17-fold enhanced PL UV emission for Al-coated ZnO with the highest carrier density ($2 \times 10^{17}$ cm$^{-3}$) was measured, which falls to a 12-fold increase for the lowest carrier density ($3 \times 10^{13}$ cm$^{-3}$). Correlative CL, PL and valence band photoelectron X-ray spectroscopy (VB-XPS) measurements also reveal that a metal surface coating (i) passivates competitive non-radiative surface recombination channel present in the uncoated ZnO and (ii) creates a surface depletion layer, which both contribute to the magnitude of UV enhancement factor. The observed UV enhancement induced by the metal coating is accompanied by changes to the relative intensity of the defect-related deep level (DL) emission bands at the near surface, which is attributed to surface band bending that ionizes donors involved in competitive radiative recombination centers.

**Methods**

The samples used in this study are polished *a*-plane ZnO single crystal 5 x 5 x 0.5 mm plates obtained from MTI Corporation USA, which were all carefully re-polished to an average surface roughness (RMS) of 1 nm. An identical polishing procedure was used every time to prevent any sample preparation induced variations in the specimens. A set of 5 samples was selected with carrier densities of $3.0 \times 10^{13}$ cm$^{-3}$, $1.8 \times 10^{14}$ cm$^{-3}$, $2.7 \times 10^{14}$ cm$^{-3}$, $3.1 \times 10^{16}$ cm$^{-3}$ and $2.0 \times 10^{17}$ cm$^{-3}$, which were determined using Hall probe measurements. The ZnO plates were sputter-coated with a 2 nm thick Al film. Only half of each sample was coated with Al to leave the uncoated side as a control and reference. Prior to coating, all samples were cleaned using a standard procedure consisting of the sonication for 20 minutes in acetone, then isopropyl alcohol and finally deionized water.

CL spectroscopy was performed in an FEI Quanta 200 SEM equipped with a Gatan CF302 continuous flow liquid helium cold stage. Light emitted from the sample was collected by a parabolic mirror and analyzed using an Ocean Optics QE Pro spectrometer. The parabolic mirror was also used to inject laser light focused at the sample plane and to collect PL, enabling direct spectral CL and PL measurements on spatially comparable regions of the samples. PL was excited with the 325 nm line of a Melles Griot He-Cd laser with varying powers from 0.1 to 3 mW. All luminescence spectra were corrected for the total response of the light collection system.

Depth-resolved CL spectra were collected at 3 kV, 5 kV, and 10 kV, providing probe depths of 40 nm, 100 nm and 350 nm below the surface, respectively, corresponding to the distance below the surface at which 70% of the integrated electron energy loss occurs, as determined by CASINO simulations [21]. In the depth-resolved measurements, the electron beam power was kept constant at 17.5 µW at each voltage by varying the electron beam current to provide the same number of injected electron-hole pairs at each acceleration voltage. At an acceleration voltage of 5 kV, the CL and PL excitation range can be considered to be comparable, assuming that the PL energy loss follows a Beer-Lambert relationship with depth z, being proportional to exp(-$\alpha$z), where $\alpha$ is the absorption coefficient for ZnO at 325 nm. However, it is significant to note that the CL excitation density at 5 kV is approximately three orders of magnitude larger than the laser source as the lateral diameter of the CL excitation volume (~ 0.040 μm) is considerably smaller that the PL excitation spot diameter (30 μm). Additionally, it is also important to note that although the excitation depth is similar for the PL and CL measurements at 5 kV, the energy loss profile of the laser is highest at the surface and then decays following the Beer-Lambert law. In contrast, the Monte Carlo simulations reveal that the CL in-depth energy loss profile versus kV has its maximum at a depth of approximately 20, 42 and 120 nm for electron beam energies of 3 kV, 5 kV, and 10 kV, respectively. This difference in the CL and PL in-depth energy loss arises because, unlike light, the energetic electrons must slow down via inelastic scattering mechanisms before they can strongly interact with the crystal lattice.

Hall effect measurements were carried using a conventional van der Pauw four-probe setup. VB-XPS was performed on the Soft X-ray Spectroscopy beam line at the Australian Synchrotron with an X-ray photon energy of 150 eV, which corresponds to a sampling depth of about 9 nm. A SPECS Phoibos 150 Hemispherical Analyzer was used for detection. All measurements were performed on the ZnO crystal with an intrinsic carrier density of $2.7 \times 10^{14}$ cm$^{-3}$ (labeled sample 1), while a set of samples with varying carrier densities from $1 \times 10^{13}$ to $2 \times 10^{17}$ cm$^{-3}$ were studied with PL spectroscopy with laser excitation powers ranging from 0.1 to 3 mW.

## Results

VB-XPS was performed to characterize the surface band structure of a ZnO specimen (sample 1) with and without the Al coating. The results of the VB-XPS measurements are depicted in Fig. 2 (a), showing the Zn 3d peak at 10.2 eV and peaks near the valence band maximum centered at 7 eV and 5 eV, that are attributed to hybridized Zn 4s-O 2p states and O 2p states, respectively [23–27]. Fig. 2 (b) shows an enlargement of the valence band onset shown in Fig. 2 (a) around 3.0 eV. The magnitude of the surface band bending ($V_{BB}$) can be determined using the following equation:

$$V_{BB} = E_g - E_{VF} - E_{CF}$$

Where, $E_g = 3.37$ eV is the band gap of ZnO at room temperature, $E_{VF}$ is the energy difference between the valence band maximum to the Fermi level at E = 0, which can be found from the intersection of the linear fits of the VB leading edge with the XPS background in Fig. 2 (b) and $E_{CF}$ is the energy difference between the Fermi level and the conduction band minimum, with $E_{CF} = (k_B T/e)\ln(N_C/n_e)$, where $k_B$ is the Boltzmann constant, $N_C$ the effective density of states in the conduction band ( $2.94 \times 10^{18}$ cm$^{-3}$ for the effective mass of $m_e^* = 0.24\ m_e$), e is the elementary charge and $n_e$ the bulk carrier density. Using the equation above, with the measured $E_{VF}$ values from the $n_e = 2.7 \times 10^{14}$ cm$^{-3}$, reveals an upward surface band bending of + 0.20 V for the uncoated ZnO and a downward band bending at the surface of - 0.22 V for the Al NP coated ZnO. An upward band bending in the uncoated n-type ZnO is widely attributed to gaseous oxygen species being chemisorbed onto the top face, capturing free electrons from the bulk and creating a surface depletion layer [22,25,28]. The downward band bending measured from the Al NP coated sample is expected as the work function ($\phi$) of the Al (3.65-4.20 eV) is smaller than the electron affinity ($\chi$) of n-type ZnO (4.5-4.7 eV) [29–32]. The resulting electron transfer from the metallic Al coating to the ZnO produces a surface accumulation layer, inducing downward band bending, consistent with the XPS VB measurements.

The width of the surface depletion layer (W) and the surface electric field strength are both dependent on $n_e$ and can be calculated using the following two equations [33]:

$$W = \sqrt{\frac{2\varepsilon_0 \varepsilon_r V_{BB}}{q n_e}} \quad \text{and} \quad E_s = \frac{q n_e W}{\varepsilon_0 \varepsilon_r},$$

where $\varepsilon_o$ is the vacuum permittivity = 8.85 x $10^{-14}$ C/Vcm and $\varepsilon_r$ = 8.66 the relative permittivity of ZnO. Using these equations, W and $E_S$ were determined using with $V_{BB}$ values measured using the VB-PS results for the $n_e$ = 2.7 x$10^{14}$ cm$^{-3}$ sample. These results reveal a similar space charge layer width, W, of 840 nm and 880 nm and a corresponding electric field at the surface, $E_S$, of 2.3 x $10^3$ Vcm$^{-1}$ (out of the sample) and 5.0 x $10^3$ Vcm$^{-1}$ (into the sample) before and after coating, respectively. It is noteworthy that the electrostatic properties of the surface depletion layer of the uncoated and Al NP coated ZnO films are similar except for the opposite polarity of the surface electric field.

Fig. 3 (a) and (b) show the results of depth-resolved CL at 3 kV, 5 kV and 10 kV measured at 80 K, where the spectra with full and dashed lines are measured from the uncoated and Al coated sides of sample 1, respectively. The spectra consist of a UV emission around 3.37 eV from free exciton (FX) and bound exciton (BX) recombination as well as broad peak positioned at 2.1 eV related to defects and impurity centers, discussed below. The UV emission of the ZnO crystal with the Al surface coating is significantly enhanced at all excitation voltages as shown in Fig 3. (a). This UV enhancement is attributed to LSP-exciton coupling as the Al NP LSP resonance is in the UV spectral range, enabling a direct dipole-dipole energy exchange with the excitons in the ZnO. The availability of this additional fast relaxation pathway increases the excitonic spontaneous emission rate, and consequently the CL intensity [3,5,6]. The largest UV enhancement of 8.3 is observed at the lowest acceleration voltage of 3 kV where excitons are excited near the Al/ZnO interface consistent with the LSP-exciton coupling mechanism. By increasing acceleration voltages to 5 kV and then 10 kV, the depth of maximum exciton excitation shifts away from the surface and accordingly the UV enhancement factor is reduced to 3.3 and 2.3, respectively.

The Al NP surface coating also modifies the DL CL emission as shown in Fig. 3 (b). Gaussian curve fitting of the DL from the single crystal ZnO shows that the broad emission consists of two peaks: a green luminescence (GL) peak at 2.3 eV and an orange luminescence (OL) at around 2.0 eV as seen in Fig. 3 (c)-(d) at an acceleration voltage of 3 kV and 10 kV, respectively. Comparative depth-resolved CL spectroscopy measurements of samples with and without Al NP surface film reveal that the Al coating quenches the GL and conversely increases the OL. The magnitude of these changes to the GL and OL is strongest near the surface at 3 kV (~ 40 nm), becoming progressively weaker at larger depths when the kV increases from 5 kV (~ 100 nm) to 10 kV (~ 350 nm). Both of these DL peaks have been attributed to radiative recombination at shallow donor – deep acceptor complexes. Here, the deep acceptor has been reported to be either substitutional lithium ($Li_{Zn}$) or interstitial oxygen ($O_i$) for the OL [34,35] and zinc vacancy ($V_{Zn}$) related centers for the GL [36,37], while shallow donors include H, Zn and Li interstitials ($H_i$, $Zn_i$ and $Li_i$) as well as substitutional Al ($Al_{Zn}$). The Al surface coating induces a shallow donor 0/+ charge transition band to bend below the surface pinned Fermi level, ionizing a neutral shallow donor involved in the GL, switching its GL charge state, quenching the GL, and increasing the emission from the competitive deeper OL recombination channel [38,39]. This mechanism explains why the magnitude of the enhancement of the OL and quenching of the GL both decrease as the kV becomes higher, because a larger number of electron hole pairs are injected at greater depths below the depletion layer, where the shallow donor charge state switching does not occur. The main point here is that the enhancement of the OL is not related to a coupling effect, which is expected because of the large difference in energy between the OL and Al LSP resonance. Note that the PL measurements of the before and after Al surface coated ZnO produce a similar UV enhancement to those described above acquired with CL excitation, including a similar change in DL as well (not shown).

The log-log plots of the CL and PL intensity versus CL and PL excitation power from $n_e = 2.7 \times 10^{14}$ cm$^{-3}$ sample with and without an Al NP coating are shown for the PL UV NBE in Fig. 4 (a), PL DL in Fig. 4 (b), CL UV NBE in Fig. 4 (c) and for CL DL in Fig. 4 (d). Here, the integrated intensity of the UV NBE emission and the DL emission were measured between 3.00 to 3.45 eV and 1.20 to 3.00 eV, respectively. The results show that the integrated UV emission of the Al NP coated ZnO is higher than that of the uncoated side with both PL and CL excitation at all excitation powers. A linear fit of these log-log excitation power density plots provides a power law exponent, m, with $\log I_{CL/PL} = m \log (P_{ex})$. Here, m is either: m < 1 (sub-linear) due to saturation of the defect, m = 1 (linear) no saturation of the recombination channel, or m > 1 typically arises from saturation of other competing radiative and non-radiative recombination pathways. Fig. 4 (a) shows that for the UV PL emission versus laser power, where m is super-linear (1.12 ± 0.01) for the uncoated ZnO and linear (1.04 ± 0.01) for the Al-coated side. The result strongly suggests that the Al coating passivates a competitive surface related recombination channel in the uncoated ZnO that saturates with increasing excitation power. The DL in uncoated ZnO exhibits a sub-linear power law exponent in Fig. 4 (b), indicating that these DL recombination pathways saturate with increasing beam power. This result is consistent with the slow relaxation lifetime of the GL [40], which is of the order of μs and much longer than that of the UV NBE emission being typically tens or hundreds of ps [41,42]. Thus, the saturation of the slower competitive GL recombination channel with increasing excitation power leads to an increase of the significantly faster UV emission, explaining both the sub-linear power law exponent of the DL emission as well as the super-linear behavior of the UV emission of the uncoated ZnO. Conversely, m is linear for the DL emission from the Al-coated ZnO, since the Al surface coating quenches the GL due to the downward band bending as described above, leaving the OL, which exhibits a relatively faster relaxation time [40]. The CL excitation density dependence results are shown in Fig. 4 (c) and Fig. 4 (d). The CL power law exponent results are the same as the PL measurements, except for the uncoated DL CL data showing stronger saturation effects due to the higher injected electron hole pair density with CL compared to PL excitation.

The NBE enhancement factor dependence on excitation power at 10 K, 80 K and 300 K using sample 1 with and without an Al surface coating in shown in Fig. 5 (a) and 5 (b) for CL and PL, respectively. The maximum UV enhancement is found at the lowest excitation power of both the electron beam and the laser. This observation can be explained using the results in Fig. 4 (a) and Fig. 4 (c), which reveal that UV emission intensity of the uncoated ZnO increases super-linearly with increasing excitation power while on the Al-coated side it scales linearly. Consequently, the NBE enhancement factor, being the ratio of these two intensities, will decrease with increasing excitation power due to the super-linear and linear power-law exponents exhibited by the ZnO samples with and without an Al coating, respectively.

To investigate the effect of the intrinsic carrier density in ZnO on the Al coating induced UV enhancement factor, power dependent PL spectroscopy was conducted on five samples with carrier densities of $3.0 \times 10^{13}$ cm$^{-3}$, $1.8 \times 10^{14}$ cm$^{-3}$, $2.7 \times 10^{14}$ cm$^{-3}$, $3.1 \times 10^{16}$ cm$^{-3}$ and $2.0 \times 10^{17}$ cm$^{-3}$. The UV PL enhancement as a function of $n_e$ and laser power, shown in Fig. 6, reveal that the UV enhancement factor increases with increasing $n_e$. Significantly, this result also shows that the strength of the emission enhancement factor can be controlled by varying $n_e$. However, the interpretation of these data requires inspection of the UV PL intensity versus $n_e$ data for the sample with and without the Al surface coating that are shown in Fig. 7. The results from the uncoated ZnO sample show that the UV PL intensity decreases linearly with increasing $n_e$ due most likely to the formation of additional competitive exciton recombination channels involving Auger relaxation processes. This explanation is supported by the excitation power dependence results displayed in Fig. 6 that reveal that the UV enhancement factor decreases with escalating laser power and that this effect becomes more pronounced with increasing $n_e$. Conversely, the decrease in UV PL in the uncoated sample with increasing $n_e$ is not observed in the Al coated specimen as shown in Fig. 7. The surface Al coating, as shown and discussed above, can increase the UV PL by quenching competitive non-radiative surface recombination. However, this effect is independent of $n_e$ and consequently cannot explain the increase in the PL emission, indicating that other phenomena in the bulk must contribute to the measured enhancement.

The VB-XPS results shown in Fig. 2 exhibit that the Al coating creates a surface depletion layer, which can be used to explain the absence of the decrease in the PL UV emission that is observed in the uncoated sample with rising $n_e$. For the sample with the lowest $n_e$ ($2 \times 10^{13}$ cm$^{-3}$), the width of the depletion region is calculated to be ~ 3,095 nm, while for the highest $n_e$ ($1 \times 10^{17}$ cm$^{-3}$) it is considerably smaller, being ~ 44 nm, noting that ~ 70% of the PL excitation occurs within a depth of 100 nm. For these two samples, the corresponding surface electric field of Al-coated ZnO crystals is ~ $1.4 \times 10^3$ Vcm$^{-1}$ and $9.6 \times 10^4$ Vcm$^{-1}$, respectively, indicating an increase of approximately two orders of magnitude. Significantly

FX are stable in ZnO, despite the presence of these large surface electric fields. This is because of the large binding energy ($E_x$) of FX in ZnO ($E_x$ = 60 meV) and their small Bohr radius ($r_x$ = 2.3 nm) [43,44]. Using these FX properties, the estimated electric field required to dissociate an FX in ZnO is ~ 2.6 x $10^5$ Vcm$^{-1}$ (= $E_x$ / q.$r_x$) [45], which is higher than the maximum electric fields observed at the ZnO surface.

The presence of the surface depletion layer increases the measured UV emission in the Al coated ZnO via two effects. First, since there are no free carriers in the depletion region, the measured UV emission is enhanced owing to: (i) the removal of competitive Auger recombination pathways and (ii) a decrease in electron – exciton scattering, causing an increase in exciton diffusion length, increasing the strength of the FX-LSP coupling. Second, as the magnitude of the surface band bending increases as $n_e$ becomes larger deeper competitive neutral donor recombination are ionized and consequently the UV emission is increased.

Fig. 6 also reveals that as the $n_e$ increases in the n-type ZnO the UV enhancement factor becomes increasingly inversely dependent on the PL excitation power. Consequently, this reduction with power density is most pronounced with the highest $n_e$ (1 x $10^{17}$ cm$^{-3}$) sample, where the UV enhancement factor decrease from 17 to 12 with an increasing laser excitation power from 0.1 to 2.1 mW, respectively, while at the lowest $n_e$ (2 x $10^{13}$ cm$^{-3}$) the reduction of the enhancement factor is negligible. Given that the FX diffusion length in ZnO is ~200 nm [46], this behavior is likely due to the creation of additional excess carrier related non-radiative relaxation channels in the bulk with increasing $n_e$. These are most likely competitive Auger recombination mechanisms at depths below the surface depletion layer that quench the FX emission as well as limiting their diffusion towards the Al metal coating – ZnO interface.

Conclusion:

Following the deposition of a 2 nm Al surface coating on an *a*-plane n-type ZnO with an $n_e$ of $2.7 \times 10^{14}$ cm$^{-3}$, the UV PL and CL significantly increases with its maximum close to the surface. An up to 17-fold increase was observed in the PL excited UV NBE from the Al coated n-type ZnO. Depth-resolved CL analysis confirms that enhanced UV emission is due to an increase in the spontaneous emission rate due to the availability on an additional fast relaxation channel involving an LSP-exciton dipole-dipole coupling mechanism. However, correlative PL, CL and VB-XPS studies revealed that a number of other processes induced by the Al coating also contribute to the total UV emission enhancement, involving (i) the formation of a surface depletion region and (ii) quenching of surface recombination channels that are present in uncoated ZnO. The results also show that the strength of the optical emission enhancement factor can be increased and regulated by decreasing the thickness of the surface depletion layer by raising the carrier density of the ZnO. Furthermore, the presence of the surface depletion layer provides a highly likely explanation for the large span of enhancement factors reported in the literature, where their effects on the optical emission were not considered. The Al coating induced increased UV ZnO emission is accompanied by changes in the ZnO deep level emission that are attributed to surface band bending, ionizing shallow donors involved in competitive radiative recombination centers.

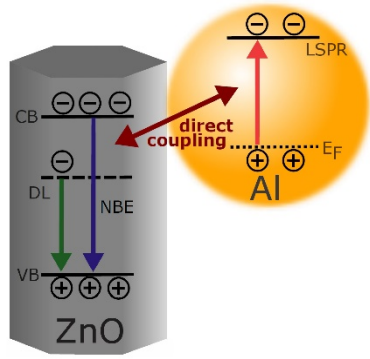

Fig. 1: Illustration of the LSP-exciton coupling mechanism.

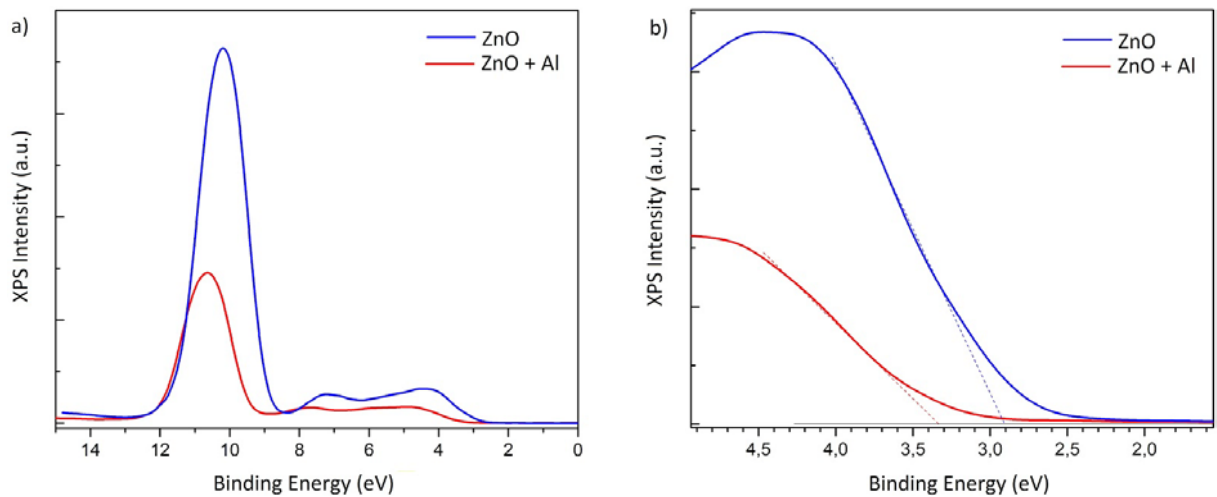

Fig. 2: (a) Valence band spectra of the uncoated (blue) and Al-coated a-plane ZnO (red) with a carrier density of $2.7 \times 10^{14}$ cm$^{-3}$. (b) Enlargement of the leading edge of valence band in (a) around 3.5 eV. The VB onset energy position is measured at the intersection of the linear fits of the VB leading edge and the background.

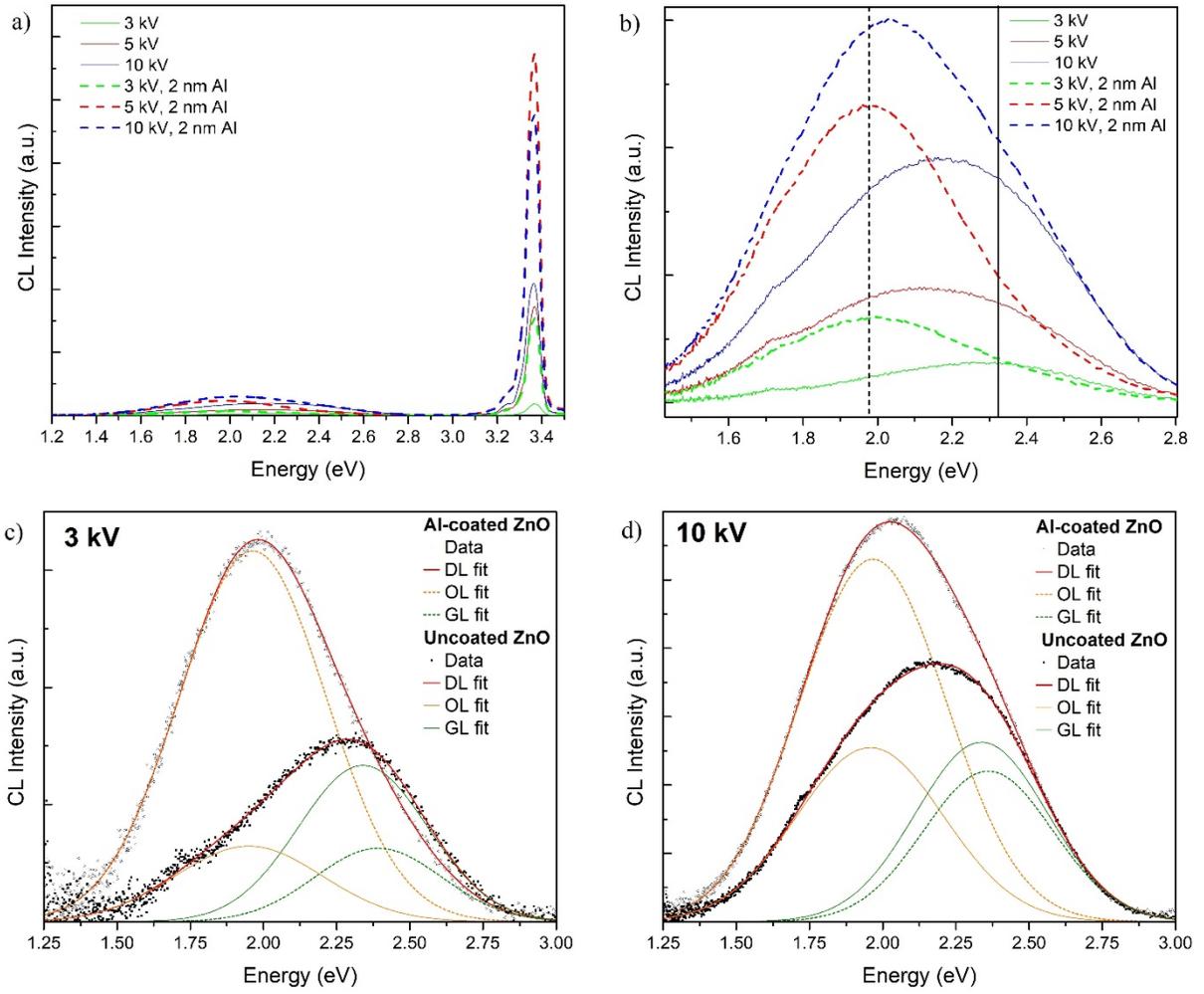

Fig. 3: (a) Depth-resolved CL spectra of uncoated (solid) and Al NP coated (dashed) *a*-plane ZnO (carrier density 2.7 x $10^{14}$ cm$^{-3}$) at T = 80 K, showing highest UV enhancement of 8.3 at accelerating voltage of 3 kV. The spectra were collected with a constant electron beam power of 17.5 μW and from a scan area of 10 μm x 10 μm. (b) Enlargement of the DL emission around 2.1 eV in (a), showing an increase in the CL emission at around 2.0 eV. (c) and (d) DL emission and fitting of the orange (OL) and green luminescence (GL) of the uncoated and Al-coated ZnO at 3 kV and 10 kV, respectively, showing a decreased GL and increased OL for the Al-coated ZnO.

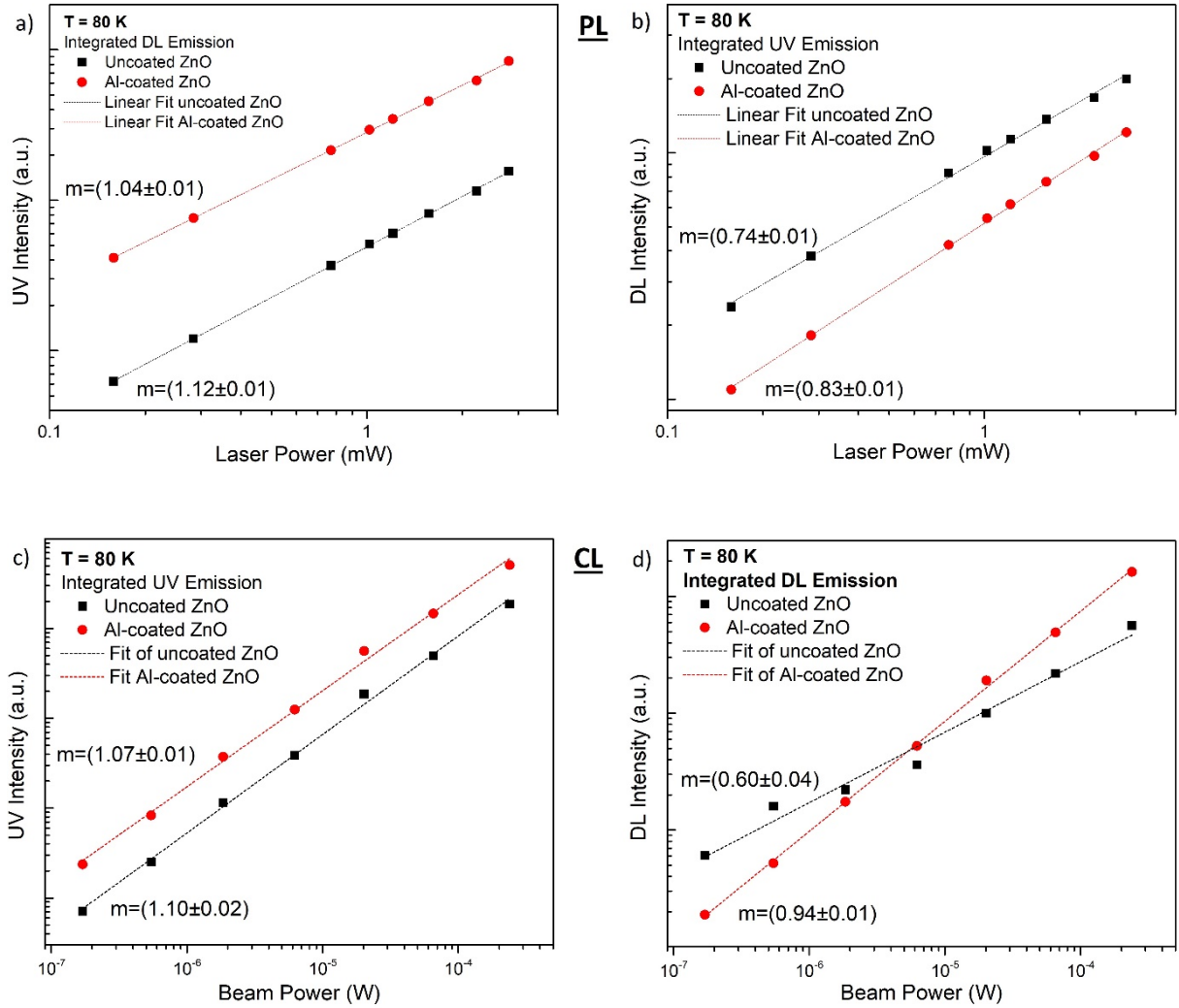

Fig. 4: Power-dependent PL of (a) the integrated NBE emission (3.00 - 3.54 eV) and (b) the DL emission (1.2 - 3.0 eV) of the uncoated (black) and Al-coated (red) ZnO with a carrier density of 2.7 x $10^{14}$ cm$^{-3}$. (c) and (d) show the corresponding power-dependent CL, graphed logarithmically. PL: excitation = 325 nm, spot size ~ 30 µm; CL: HV = 5 kV, scan area = 15 µm x 15 µm; T = 80 K.

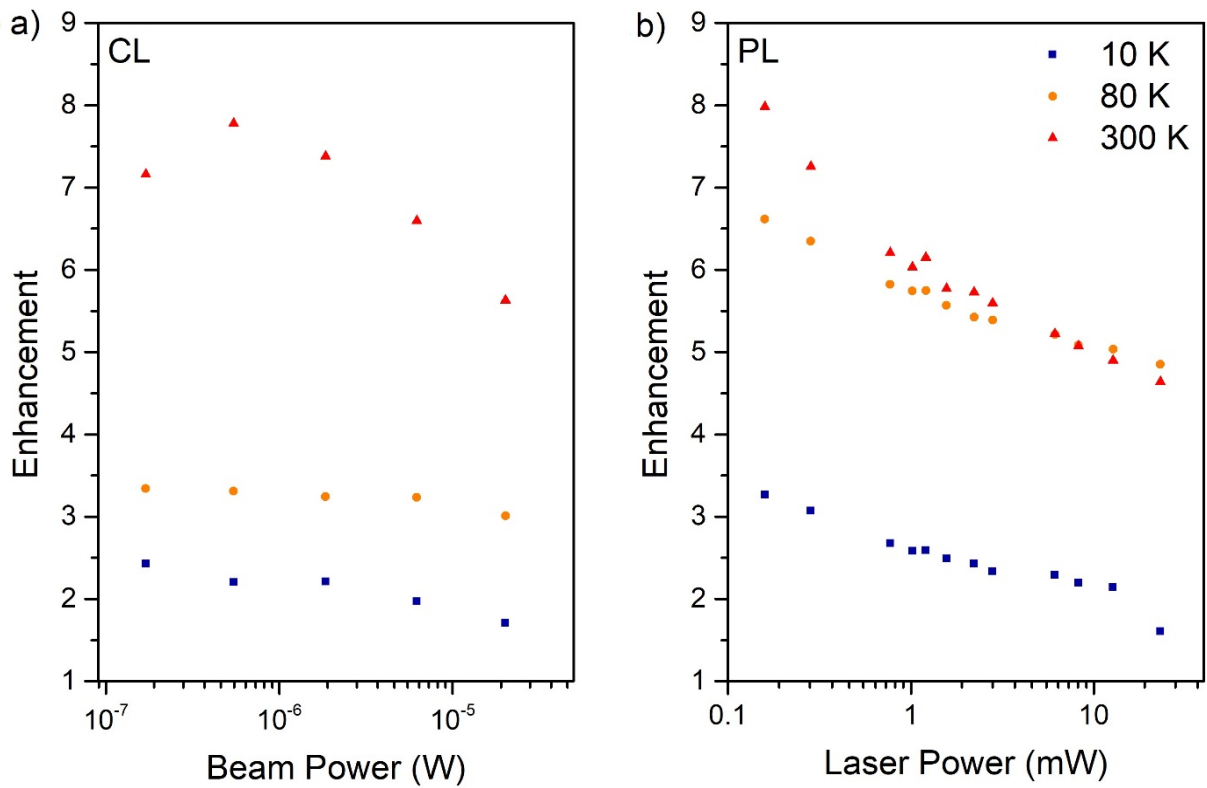

Fig. 5: (a) CL and (b) PL UV enhancement factor as a function of CL and PL excitation power from sample with an intrinsic carrier density of $2.7 \times 10^{14}$ cm$^{-3}$ at temperatures of 10, 80 and 300 K.

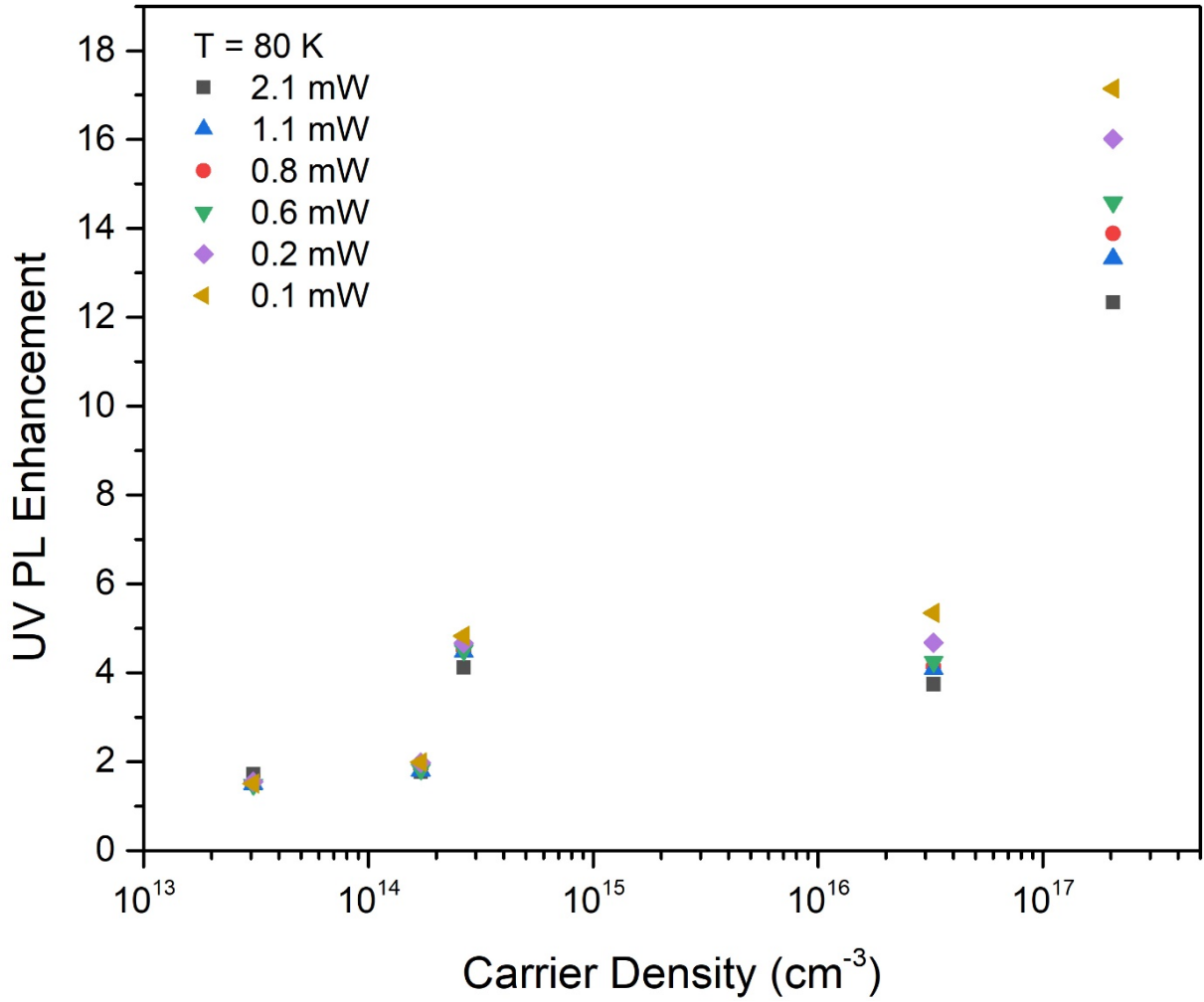

Fig. 6: UV enhancement of Al-coated *a*-plane n-type ZnO with different carrier densities ranging from $10^{13}$ to $10^{17}$ cm$^{-3}$. The laser excitation power was varied from 0.1 mW to 2.1 mW at a wavelength of 325 nm with a spot size of ~ 30 μm.

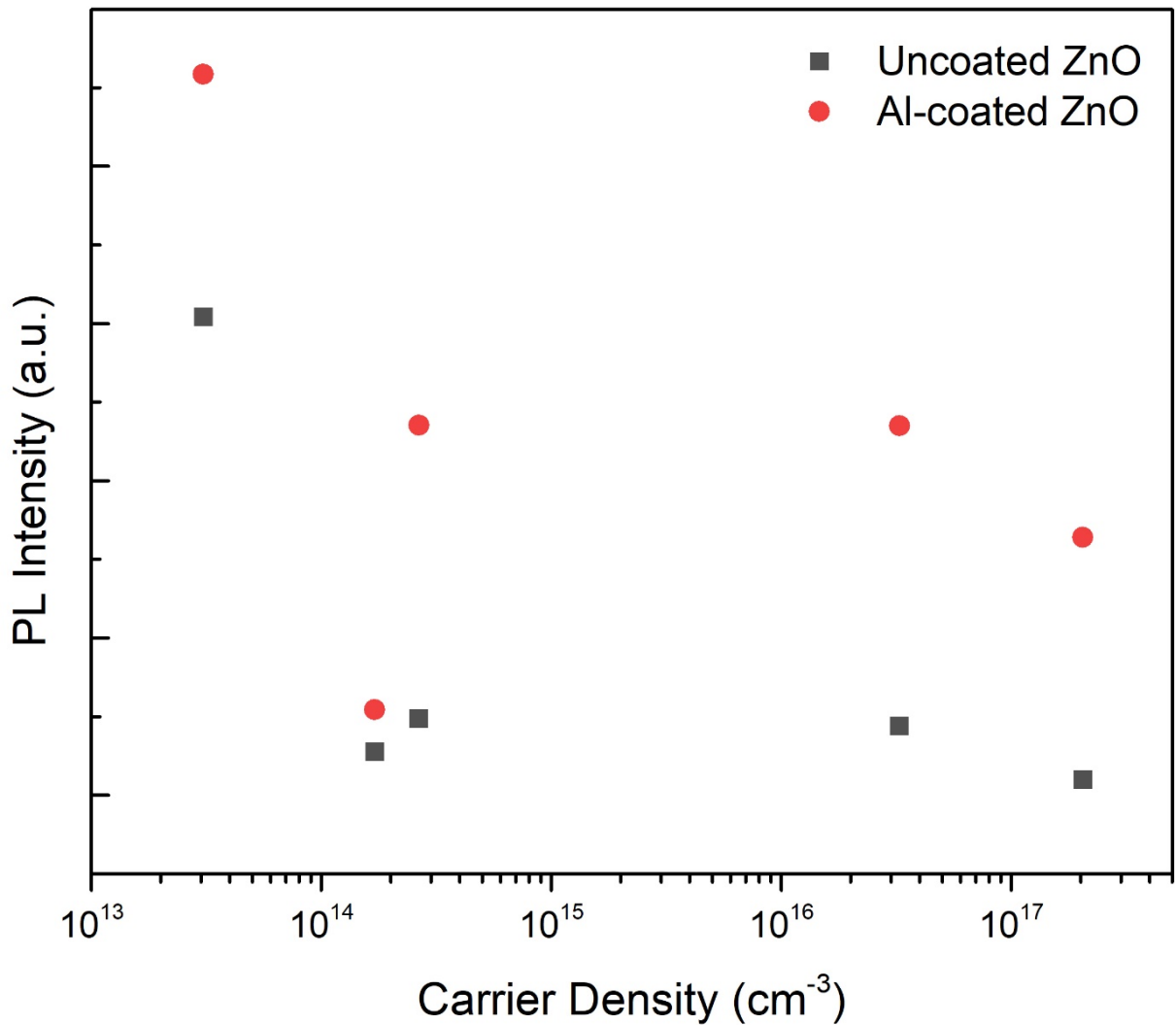

Fig. 7: PL Intensity before and after Al coating versus carrier density, ne (cm-3). The uncoated sample exhibits a linear decrease in the PL intensity with increase ne, which is absent in the coated sample due to the presence of a surface depletion layer, where its thickness decreases with increasing ne.

References.

Acknowledgement:

The authors would like to thank Katie McBean, Geoff McCredie and Dr. Angus Gentle for the technical support at UTS, and Dr. Marie Wintrebert-Fouquet for her assistance with the Hall effect measurements at BluGlass Australia.
This research work was financially supported by the Australian Research Council, Discovery Program Scheme, Discovery Project DP150103317.